\documentclass{ws-procs9x6-cpt22}
\begin{document}

\newcommand{\refeq}[1]{(\ref{#1})}
\def\etal {{\it et al.}}

\title{The ADM Formulation of the SME Gravity}

\author{Carlos M. Reyes }

\address{Centro de Ciencias Exactas, Universidad del B\'{i}o-B\'{i}o \\
Avda.~Andr\'es Bello 720, Chill\'{a}n, 3800708, Chile}

\begin{abstract}
The Hamiltonian formulation of the gravitational sector of the Standard-Model Extension (SME)
with nondynamical fields $u$ and $s^{\mu \nu}$ is studied. 
We provide the relevant Hamiltonians that describe the constrained phase 
space and the dynamics of the induced metric on the
ADM hypersurface.
The generalization of the Gibbons-Hawking-York boundary term
has been crucial to preventing
 second time-derivatives of the metric tensor in the Hamiltonians. 
By extracting the dynamics and constraints from the Einstein equations we
have proved the equivalence between the Lagrangian and Hamiltonian formulations.
\end{abstract}

\bodymatter
\section{Introduction}
Einstein's theory of General Relativity (GR) has profoundly shaped our understanding of the physical world at large and small scales. 
GR tells us that
spacetime is a four-dimensional manifold on which a metric 
solving Einstein's equations defines how matter moves,
and matter can deform the spacetime geometry.

Soon after the advent of Quantum Field Theory  
and the success of the perturbative method for the Standard Model (SM)
of particles, it was explored whether the perturbative techniques could be applied on
the gravitational field.
However, it was shown that pure gravity
has nonrenormalizable divergences at two loops\cite{Goroff:1985sz} and becomes
more singular when coupling to matter, in particular for the electromagnetic\cite{Deser:1974zzd} and Dirac\cite{Deser:1974cy} fields. 

The 
incompatibility of SM and GR 
suggest that gravity is a low-energy approximation of a fundamental theory. 
New physics in the form of Lorentz and CPT symmetry breaking has been postulated as one possible effect that could be detected at low energies.
The Standard-Model Extension
(SME) is a comprehensive effective-field theory framework developed to study 
departures from CPT and Lorentz symmetry in particle physics\cite{Colladay:1996iz,Colladay2} and diffeomorphism and local Lorentz violations in gravity.\cite{Kostelecky:2003fs}
\section{Hamiltonian form of the SME gravity}
We start with the gravitational action\cite{Kostelecky:2003fs} without a cosmological constant
\begin{equation}
\label{eq:minimal-gravity-sme-reformulated}
S=\int_{\mathcal{M}}\mathrm{d}^4x\frac{\sqrt{-g}}{2\kappa}\left[(1-u){}^{(4)}R+s^{\mu\nu}{}^{(4)}R_{\mu\nu}\right]\,,
\end{equation}
with $\kappa=8\pi G_N$, $G_N$ the Newton constant, the Ricci tensor ${}^{(4)}R_{\mu\nu}$ 
and the associated Ricci scalar ${}^{(4)}R:={}^{(4)}R^{\mu}_{\phantom{\mu}\mu}$ of the 
four-dimensional spacetime manifold $\mathcal{M}$ with metric tensor $g_{\mu\nu}$ and $g:=\det(g_{\mu\nu})$. 
Furthermore, $u=u(x)$ and $s^{\mu\nu}=s^{\mu\nu}(x)$ are nondynamical background fields having a generic spacetime dependence. 

The modified Einstein equations of motion for the Lagrangian~\eqref{eq:minimal-gravity-sme-reformulated} are
\begin{align}
\label{eq:einstein-equations-modified-generic}
0&=(1-u){}^{(4)}G^{\mu\nu}+\frac{1}{2}(\nabla^{\mu}\nabla^{\nu}u
+\nabla^{\nu}\nabla^{\mu}u)-g^{\mu\nu}\square u    -\frac{1}{2}\left(s^{\alpha\beta}{}^{(4)}R_{\alpha\beta}g^{\mu\nu}
\right. \nonumber  \\ & \phantom{{}={}}\hspace{1.5cm} \left.  +\nabla_{\alpha}\nabla^{\mu}
s^{\alpha\nu}+\nabla_{\alpha}\nabla^{\nu}s^{\alpha\mu}-\square s^{\mu\nu}  -g^{\mu\nu}\nabla_{\alpha}\nabla_{\beta}s^{\alpha\beta}\right)\,,
\end{align}
where $G^{\mu \nu}$ is the Einstein tensor, $\Box=\nabla_{\mu} \nabla^{\mu}$, $\nabla_{\mu}$ is the covariant derivative
 and we have used that the backgrounds have zero fluctuations.

The Hamiltonian formulation by definitions needs a time variable and, hence, our next step is to decompose spacetime.
We consider the $3+1$ 
decomposition of spacetime due to Arnowitt, Deser and Misner (ADM).\cite{ADM_formulation}
The ADM decomposition of the metric tensor 
$g_{\mu \nu}$ turns to be
 $g_{00}=-N^2+q^{ij}N_iN_j$, $g_{0i}=N_i$ and $g_{ij}=q_{ij}$,
where 
$q_{ij}$ is the induced metric on the hypersurface $\Sigma_t$, and $N$, $N^i$ are called the \emph{lapse} and the \emph{shift}, respectively.

The Legendre transformation of the Lagrangian~\eqref{eq:minimal-gravity-sme-reformulated}
leads to
the canonical Hamiltonians described in Ref.\ \refcite{ADM}.
To give a general idea, we focus on the derivation of the Hamiltonian
in the $u$ sector. One finds the conjugate momentum 
\begin{align}
\pi^{ij}&=\frac{\sqrt{q}}{2\kappa}\left[(1-u)(K^{ij}-q^{ij}K)+\frac{1}{N}q^{ij}\mathcal{L}_mu\right]  \,,
\end{align}
and the Hamiltonian
\begin{align}
{H_u}&=\int_{\Sigma_t}\mathrm{d}^3x\,\left[-\frac{\sqrt{q}}{2\kappa}N\left((1-u)R+2D^iD_iu\right)+\frac{\mathcal{L}_mu}{1-u}\left(\pi-\frac{3}{4}\frac{\sqrt{q}}{\kappa N}\mathcal{L}_mu\right)   \right. \nonumber  \\  &\phantom{{}={}}\hspace{1.5cm} \left.  +\frac{2\kappa N}{\sqrt{q}(1-u)}\left(\pi^{ij}\pi_{ij}-\frac{\pi^2}{2}\right)  -2(D_i\pi^{ij})N_j\right]\,.
\end{align}
The Hamiltonians in the $s^{\mu \nu}$ sector follows by considering
the ADM decomposition of the background field $s^{\mu\nu}$, which is
\begin{equation}
\label{eq:decomposition-s}
s^{\alpha\beta}=q^{\alpha}_{\phantom{\alpha}\mu}q^{\beta}_{\phantom{\beta}\nu}s^{\mu\nu}-(q^{\alpha}_{\phantom{\alpha}\nu}n^{\beta}
+q^{\beta}_{\phantom{\beta}\nu}n^{\alpha})s^{\nu \mathbf{n}}+n^{\alpha}n^{\beta}s^{\mathbf{nn}}\,,
\end{equation}
where $q^{\mu}_{\phantom{\mu}\nu}$ projects a tensor or a part of it into $\Sigma_t$. We have three more sectors the $s^{ij}:=q^i_{\phantom{i}\mu}q^j_{\phantom{j}\nu}s^{\mu\nu}$ as the purely spacelike sector of $s^{\mu\nu}$ that lives in $\Sigma_t$ entirely, the $s^{i\mathbf{n}}:=q^i_{\phantom{i}\mu}n_{\nu}s^{\mu\nu}$ be the vector-valued piece and $s^{\mathbf{n}\mathbf{n}}:=n_{\mu}n_{\nu}s^{\mu\nu}$ the scalar part.
The details can be found in Ref.\ \refcite{ADM,ADM_Dynamics} and a similar treatment in Ref.\ \refcite{Alt_Hamiltonian_SME}.
\section{Dynamics and constraints}
To find the Hamiltonians we have introduced the extended Gibbons-Hawking-York term
\begin{equation}
\label{eq:modified-GHY}
S_{\substack{\text{ext} \\ \text{GHY}}}=\frac{\varepsilon}{2\kappa}\oint_{\partial\mathcal{M}} \mathrm{d}^3y\,\sqrt{q}\,\left[2(1-u)K-s^{\mathbf{nn}}K+K_{ij}s^{ij}\right]\,,
\end{equation}
where the parameter $\varepsilon=\mp 1$ for a spacelike (timelike) boundary $\partial\mathcal{M}$ of
 the spacetime manifold $\mathcal{M}$, $K_{ij}$ is the extrinsic curvature, $q^{ij}K_{ij}=K$ its trace and the integral runs over the coordinates $y^i$ defined on this boundary. 
In Ref.\ \refcite{ADM_Dynamics}, we introduce a second
 boundary term for $u$ and $s^{\mathbf{nn}}$ that is of plainly different nature compared to that of~\eqref{eq:modified-GHY}
\begin{equation}
\label{eq:additional-boundary-term}
S_{\partial\Sigma}=-\frac{1}{2\kappa}\oint_{\partial\Sigma_t} \mathrm{d}^2z\,\sqrt{q}r_l\left[ND^l(2u+s^{\mathbf{nn}})\right]\,.
\end{equation}
Consider the equation of motion in the $u$ sector 
\begin{align}
\label{eq:modified-einstein-u}
Q^{\mu\nu}:=(1-u){}^{(4)}G^{\mu\nu}+\nabla^{\mu}\nabla^{\nu}u-g^{\mu\nu}\square u=0\,.
\end{align}
Now, the projected equation of motion into $\Sigma_t$ is
\begin{align}
(\vec{\boldsymbol{q}}^{*}\mathbf{Q})^{ij}&=\frac{2\kappa}{N\sqrt{q}}\dot{\pi}^{ij}+(1-u)
\left(R^{ij}-\frac{R}{2}q^{ij}\right)+\frac{1}{N}\left(q^{ij}
D_kD^k[(1-u)N]  \right. \notag \\   & \left. -D^iD^j[(1-u)N]\right) +q^{ij}a^kD_ku-(a^iD^ju+a^jD^iu) 
+\frac{4\kappa^2}{q(1-u)}  \notag  \\  &\times
   \left[2\pi^{ik}\pi_k^{\phantom{k}j}-\pi\pi^{ij}    
 -\frac{1}{2}\left(\pi_{kl}\pi^{kl}-\frac{\pi^2}{2}\right)q^{ij}\right]  +\frac{\mathcal{L}_mu}{N(1-u)} \notag  \\ &\times \left(\frac{2\kappa}
{\sqrt{q}}\pi^{ij}-\frac{3}{4N}q^{ij}\mathcal{L}_mu\right)\,,
\end{align}
the purely orthogonal projection given by
\begin{align}
2\mathbf{Q}(\mathbf{n},\mathbf{n})&=(1-u)R+2D_iD^iu-\frac{4\kappa^2}{q(1-u)}\left(\pi_{ij}\pi^{ij}-\frac{\pi^2}{2}\right)\notag  \\ & -\frac{3(\mathcal{L}_mu)^2}{2(1-u)N^2}\,,
\end{align}
and the mixed projection by
\begin{align}
2\mathbf{Q}^k(\vec{\boldsymbol{q}}(.),\mathbf{n})&=\frac{2\kappa}{\sqrt{q}}\left[2D_i\pi^{ik}+\frac{\pi}{1-u}D^ku\right]-\frac{3\mathcal{L}_mu}{(1-u)N}D^ku\,.
\end{align}
We have defined
\begin{equation}
\mathbf{T}(\mathbf{n},\mathbf{n}):=n_{\mu}n_{\nu}T^{\mu\nu}\,,\quad \mathbf{T}^{\varrho}(\vec{\boldsymbol{q}}(.),\mathbf{n}):=q^{\varrho}_{\phantom{\varrho}\mu}n_{\nu}T^{\mu\nu}\,,
\end{equation}
and
\begin{align}
(\vec{\boldsymbol{q}}^{*}\mathbf{T})^{\alpha_1\dots\alpha_s}_{\phantom{\alpha_1\dots\alpha_s}\beta_1\dots\beta_t}&:=q^{\alpha_1}_{\phantom{\alpha_1}\gamma_1}\dots q^{\alpha_s}_{\phantom{\alpha_s}\gamma_s}  q^{\delta_1}_{\phantom{\delta_1}\beta_1}\dots q^{\delta_t}_{\phantom{\delta_t}\beta_t}T^{\gamma_1\dots\gamma_s}_{\phantom{\gamma_1\dots\gamma_s}\delta_1\dots\delta_t}\,.
\end{align}
In Ref.\ \refcite{ADM_Dynamics} we have shown the equivalence 
between the Lagrangian and Hamiltonian formulations.
\section{Conclusions}
We have performed the ADM decomposition of the 
 gravitational sector of the SME.
A crucial part of the derivation was to extend the Hawking-Gibbons-York boundary term in order to avoid second 
time-derivatives on the metric in the Hamiltonians. 
We have extracted the dynamics and constraints from the Einstein equation and compared with those obtained in the Hamiltonian formulation. 
In addition, the ADM formalism
has stimulated the works on cosmological applications.\cite{ADM_Cosmology,Nilsson:2022mzq}
\section*{Acknowledgments}
This work was supported by Fondecyt Regular grant No. 1191553, Chile.
I want to thank Yuri Bonder for many useful discussions.

\end{document}